\documentclass[aps,prd,preprintnumbers,nofootinbib,twocolumn]{revtex4}
\usepackage{bm}
\usepackage{latexsym}
\usepackage{dcolumn}
\usepackage{amsmath,amsfonts,amssymb}
\usepackage{graphicx,epsfig}
\usepackage{psfrag}
\usepackage{amsthm}

\def\be {\begin{equation}}
\def\ee {\end{equation}}
\def\bea {\begin{eqnarray}}
\def\eea {\end{eqnarray}}
\def\bc {\begin{center}}
\def\ec {\end{center}}
\def\bfg {\begin{figure}}
\def\efg {\end{figure}}
\def\bi {\begin{itemize}}
\def\ei {\end{itemize}}

\def\la {\label}
\def\le {\left}
\def\ri {\right}
\def\pa {\partial}

\def\no {\noindent}

\def\vs {\vspace}

\def\a  {\alpha}

\def\beq{\begin{equation}}
\def\eeq{\end{equation}}
\def\br{\begin{eqnarray}}
\def\er{\end{eqnarray}}
\newcommand{\eel}[1] {\label{#1}\end{equation}}

\newcommand{\bdm}{\begin{displaymath}}
\newcommand{\edm}{\end{displaymath}}

\begin{document}
\title{Cosmic coincidence or graviton mass?
}

\author{Saurya Das} \email[email: ]{saurya.das@uleth.ca}

\affiliation{Department of Physics and Astronomy,
University of Lethbridge, 4401 University Drive,
Lethbridge, Alberta, Canada T1K 3M4 \\
{}\\
{\bf This essay received Honorable Mention in 
the 2014 Gravity Research Foundation Essay Competition \\
}
}

\begin{abstract}
Using the quantum corrected Friedmann equation, obtained from the quantum Raychudhuri 
equation, and assuming a small mass of the graviton (but consistent with observations and theory),
we propose a resolution of the smallness prroblem 
(why is observed vacuum energy so small?) and the coincidence problem 
(why does it constitute most of the universe, about 70\%, in the current epoch?).
\end{abstract}

\maketitle


%
%
As a result of some remarkable recent advances in astrophysical observations of high red-shift
supernovae \cite{perlmutter,riess} and the cosmic microwave background radiation \cite{wmap,planck}, 
we now have a better than ever understanding of the large scale structure of our universe and its evolution. 
It is now more or less accepted that it was created at the big-bang singularity about $14$ billion years ago, that it underwent a short but rapid inflationary phase, then an expanding phase in which it transited from radiation to a matter dominated era, and is currently homogeneous and isotropic and in an accelerating phase
and made up of about $70\%$ {\it Dark Matter} characterized by a pressure to density ratio $w\equiv p/\rho=-1$, and the remaining non-relativistic matter (mostly dark), with $w=0$.
One also assumes that that universe obeys the laws of general relativity and quantum mechanics, the latter being important at very early times.
Beneath this apparently simplicity problems remain however, among them perhaps the most notorious
being the extremely small value of the cosmological constant $\Lambda$ for it to be a candidate for dark energy, about $10^{-124}$ in Planck units, known as the {\it smallness problem}
(e.g. vacuum energy of quantum fields predict $50$ orders of magnitude or more, greater than the observed value), and also its almost equality with $H_0^2/c^2$, where $H_0=$ the current value of the Hubble parameter, also known as the {\it coincidence problem}. In this article, we show that both these problems can be resolved in one stroke, provided one assumes that the origin of $\Lambda$ lies in the quantum wavefunction of gravitons (of photons) which pervade our universe, albeit having a small mass, but consistent with all observations.

Since the Friedmann equation, the guiding equation of cosmology, can be derived
from the Raychaudhuri equation,
we start with the recently obtained quantum corrected Raychaudhuri equation (QRE), obtained by replacing geodesics with quantal (Bohmian) trajectories \cite{bohm}, associated with
a wavefunction $\psi={\cal R}e^{iS}$ of the fluid or condensate filling our universe
(${\cal R} (x^\a),S (x^a)=$ real), and giving rise to
the four velocity field $u_a = (\hbar/m) \pa_a S$, and expansion
$\theta=Tr(u_{a;b}) = h^{ab} u_{a;b},~h_{ab}=g_{ab}-u_au_b$
\cite{sd}
\footnote{We use the metric signature $(-,+,+,+)$ here, as opposed to $(+,-,-,-)$ in \cite{sd},
resulting in opposite sign of the $\hbar^2$ terms. Here we concentrate
on the more important of the two correction terms. }.
\bea
\frac{d\theta}{d\lambda} =&& - \frac{1}{3}~\theta^2
- R_{cd} u^c u^d
+  \frac{\hbar^2}{m^2} h^{ab} \le( \frac{\Box {\cal R}}{\cal R} \ri)_{;a;b}
\la{qre2a} 
\eea
The second order Friedmann equation satisfied by
the scale factor $a(t)$ can be derived from the above, by replacing
$\theta = 3{\dot a}/{a}$ 
and
$R_{cd} u^c u^d \rightarrow \frac{4\pi G}{3} (\rho+3p) 
$ \cite{akrbook}
\bea
\frac{\ddot a}{a} &&= - \frac{4\pi G}{3} \le( \rho + 3p \ri) 
+ \frac{\hbar^2}{3 m^2} h^{ab} \le( \frac{\Box {\cal R}}{\cal R} \ri)_{;a;b}~.
\la{frw1}
\eea
The quantum correction (${\cal O}(\hbar^2)$) term in  Eqs.(\ref{qre2a}) and (\ref{frw1}),
also known as {\it `quantum potential'} (a term coined by Bohm \cite{bohm}),
vanish in the $\hbar\rightarrow 0$ limit giving back
the classical Raychaudhuri and the Friedmann equations.
Note that since Bohmian trajectories do not cross
\cite{holland,nocrossing},
it follows that even when $\theta$ (or $\dot a$) $\rightarrow -\infty$, the actual trajectories
(as opposed to geodesics) do not converge and there is no counterpart of
geodesic incompleteness or the classical singularity theorems,
and singularities such as big bang or big crunch can in fact avoided.
Next, we interpret the correction term as the cosmological constant
\bea
\Lambda_Q = \frac{\hbar^2}{m^2 c^2} h^{ab} \le( \frac{\Box {\cal R}}{\cal R} \ri)_{;a;b}~.
\la{qlambda}
\eea
Although $\Lambda_Q$ depends on the form of the amplitude of the
wavefunction ${\cal R}$, for any reasonable form
such as a Gaussian wave packet $\psi \sim \exp(-r^2/L^2)$, or for one which results when an
interaction is included in the scalar field equation $[\Box + g|\psi|^2 -k ]\psi=0$, namely
$\psi=\psi_0\tanh(r/L\sqrt{2}) ~(g>0)$ and $\psi=\sqrt{2}~\psi_0~{\mbox {sech}}(r/L)~(g<0)$
\cite{rogel} it can be easily shown that $(\Box {\cal R}/{\cal R})_{;a;b} \approx 1/L^4$,
where $L$ is the characteristic length scale in the problem, typically the
Compton wavelength $L=\hbar/mc$ \cite{wachter} over which the wavefunction is non-vanishing.
This gives
\bea
\Lambda_Q = \frac{1}{L^2} =\le( \frac{m c}{\hbar} \ri)^2 ~ \la{lambda1}
\eea
which has the correct sign as the observed cosmological constant.
Next to estimate its magnitude,
we note that if $L$ is identified with the linear dimension of our observable universe,
then $m$ can be regarded as the small mass of gravitons (or photons),
with gravity (or Coulomb field) following a Yukawa type of force law
$
F = - \frac{Gm_1 m_2}{r^2} \exp(-r/L)~.
$
Since gravity (and light) has not been tested beyond the above length scale, this interpretation is
natural and may in fact be unavoidable.
If one invokes periodic boundary conditions, this is also the mass of the lowest
Kaluza-Klein modes.
Substituting $L = 1.4 \times 10^{26}$ {\it metre}, one obtains
$m \approx 10^{-68}~kg$ or $10^{-32}~eV$,
quite consistent with the estimated bounds on graviton masses from various experiments \cite{graviton},
and also from theoretical considerations \cite{zwicky,mann,derham,majid}.
%
Finally, plugging in the above value of $L$ in Eq.(\ref{lambda1}), we get
\bea
\Lambda_Q && = 10^{-52}~\text{(metre)}^{-2} \\
&& = 10^{-123}~~(\text{in Planck units})~,
\eea
which is indeed the observed value.
Also since the size of the observable universe is about $c/H_0$, where
$H_0$ is the current value of the Hubble parameter, one sees why
the above value of $\Lambda_Q$ numerically equals $H_0^2/c^2$
(which is $8\pi G/3 c^4 \times \rho_{crit}$, the critical density), offering a viable
explanation of the coincidence problem.

In summary, gravitons and photons which pervade our universe 
and collectively described by a wavefunction, necessarily
give rise to a quantum potential which manifests as cosmological
constant in the quantum corrected Friedmann equation. Furthermore 
for all reasonable choices of the wavefunction, 
its magnitude turns out to be $(\mbox{\it observable universe size})^{-2}$, which
remarkably matches with the accepted minute value of the cosmological constant.
Note that the argument goes beyond just a dimensional one, and
this identification (again which appears unavoidable) readily provides a natural explanation of the 
smallness and coincidence problem in cosmology, which were sought for a long time. 
Further extensions of these results can be found in \cite{alidas}.

%
%

\vs{.2cm}
\no {\bf Acknowledgment}

\no
This work is supported by the Natural Sciences and Engineering
Research Council of Canada.




\begin{thebibliography}{99}



\bibitem{perlmutter} S. Perlmutter et al.
Astrophysical J. {\bf 517 (2)} (1999) 565�86 [arXiv:astro-ph/9812133].

\bibitem{riess} A. G. Riess et al., Astron. J. {\bf 116} (1998) 1009 [arXiv:astro-ph/9805201].

\bibitem{wmap} G. Hinshaw et al, arXiv:1212.5226 [astro-ph.CO].

\bibitem{planck} P. A. R. Ade et al, arXiv:1303.5062.

\bibitem{bohm}
 D. Bohm, Phys. Rev. {\bf 85} (1952) 166; D. Bohm, B. J.
 Hiley, P. N. Kaloyerou, Phys. Rep. {\bf 144}, No.6 (1987) 321.

\bibitem{sd} S. Das, Phys. Rev. {\bf D89} (2014) 084068 [arXiv:1311.6539].

\bibitem{akrbook}
A. K. Raychaudhuri, {\it Theoretical Cosmology}, Oxford (1979).

\bibitem{holland}
P. R. Holland, {\it The Quantum Theory of Motion}, Cambridge (1993).

\bibitem{nocrossing}
C. Phillipidis, C. Dewdney, B. J. Hiley, Nuovo Cimento,
{\bf 52B} (1979) 15-28;
D. A. Deckert, D. D\"urr, P. Pickl, J. Phys. Chem. A, 111,
41 (2007) 10325; A. S. Sanz, J. Phys.: Conf. Ser. {\bf 361},
012016 (2012);
%
A. Figalli, C. Klein, P. Markowich, C. Sparber,
{\it WKB analysis of Bohmian dynamics}, arXiv:1202.3134.


\bibitem{rogel} J. Rogel-Salazar, Eur. J. Phys. {\bf 34} (2013) 247 [arXiv:1301.2073];
N. \"Uzar, S. Deniz Han, T. T\"ufekci, E. Aydiner, arXiv: 1203.3352.

\bibitem{wachter} A. Wachter, {\it Relativistic Quantum Mechanics}, Springer (2010).

\bibitem{graviton}
C. M. Will, Phys. Rev. {\bf D57} (1998) 2061;
L. S. Finn, P. J. Sutton Phys. Rev. {\bf D65} (2002) 044022 [arXiv:gr-qc/0109049];
A. S. Goldhaber, M. M. Nieto, Rev. Mod. Phys. {\bf 82} (2010) 939 [arXiv:0809.1003]
E. Berti, J, Gair, A. Sesana, Phys. Rev. {\bf D84} (2011) 101501(R).

\bibitem{zwicky} F. Zwicky,
{\it Cosmic and terretrial tests for the rest mass of gravitons},
Publications of the Astronomical Society of the Pacific, Vol. 73, No. 434, p.314.

\bibitem{mann} J. R. Mureika, R. B. Mann,
Mod. Phys. Lett. {\bf A26}  (2011) 171-181 [arXiv:1005.2214].

\bibitem{derham} C. de Rham, G. Gabadadze, L. Heisenberg, D. Pirtskhalava,
Phys. Rev. {\bf D83} (2011) 103516 [arXiv:1010.1780].

\bibitem{majid} S. Majid, arXiv:1401.0673.

\bibitem{alidas} A. F. Ali, S. Das, 
{\it Cosmology from quantum potential}, arXiv:1404.3093. 


\end{thebibliography}
\end{document}